# Improved Majority Identification by the Coarsened Majority Automaton

**Authors**

David Peak, Charles G. Torre, and Jenny R. Whiteley
Physics Department, Utah State University, Logan, Utah 84322

**Abstract**

The "initial majority identification task" is a fundamental test problem in cellular automaton research. To pass the test, an automaton must evolve to a uniform configuration consisting of the state that was in the majority for any initial configuration, employing only its internal, local dynamics. It is known that no two-state automaton can perform the majority task perfectly. Thus, it is a matter of continuing interest to identify and analyze new automata with improved majority identification capability. Here, we show that a "coarsened" version of one of the best majority identifiers can out-perform its "parent" automaton while simultaneously reducing the associated computational costs.

**Introduction**

A 2011 paper with the provocative title, "Uninformed Individuals Promote Democratic Consensus in Animal Groups" [Couzin (1)], reports experiments in which minnows initially trained to school toward either a yellow or a blue target are *mixed together*. As long as there is only a small initial majority the two subpopulations continue to school independently toward the target each was trained to prefer. When a large number of *untrained* fish is added, however, frequently the entire new population is found to school toward the target favored by the initial small majority (thus "promoting democratic consensus"). This is a surprising result, given that adding the untrained fish causes the initial net target preference of the entire enlarged population to be even smaller than before the addition.

In the succeeding years, hundreds of papers have referred this result. One of these ("Undecided Cliques Promote Consensus in the Directed Majority Automaton" [Christensen (2)]) focuses on the minimal ingredients needed for the fish to come to consensus. The paper assumes there are no leader fish responsible for coordinating the population's behaviors; consensus, it supposes, emerges via collective interactions. The authors postulate that the minnows employed in the experiment are in essence engaged in solving the *2d cellular automaton "majority identification task."*

In the "Undecided Cliques" paper, each fish is modelled as looking at a small number of neighbors in the direction of the target it currently prefers. As a result, it either keeps that preference or switches, depending on the majority preference of only its local group at that moment. In the experiment and in the model, fish on the outer edges of a school often have too few neighbors in their preferred direction to implement this rule; in that case, in the model at least, they randomly "guess" whether to switch or not. In this model the initial majority is identified if, after a prescribed time, all (or almost all) of the fish are oriented toward the correct target. The model naturally and correctly explains (a) how adding "uninformed" fish can "promote consensus" and (b) why larger (and therefore typically more compact) populations are better at performing the task.

While these model results are interesting and might potentially be useful for providing an alternative interpretation of the observations in [Couzin (1), Katz (3)], they don't clarify the fundamental question of how the Directed Majority Automaton (DMA) actually performs majority identification, nor how, if possible, it might be improved. To probe these issues, we study here a "no-guessing" version of the DMA in which the space of the automaton has no edges, and updates are deterministic. We propose that renormalization and universality arguments yield important insight into how the DMA works. In particular, we show that majority identification by the DMA is equivalent to a *directed percolation phase transition*. Using these arguments, we develop a coarsened version of the DMA that performs the majority identification task substantially better and also much more rapidly.

**Definition and Performance of the Directed Majority Automaton**

The space of our DMA is an $N$x$N$ array of cells with toroidal boundary conditions. (Note that in the majority identification literature, the deterministic DMA we define here is also referred to as the "2dGKL automaton" [Messinger (4), Cenek (5)].) Each cell can have one of two states—for example, +/–1. In each time step, each cell executes the "NE/SW" rule: if at time $t$ the cell's state is +1, then at time $t + 1$ the cell's state is the current majority of itself and its two nearest neighbors to the north and east; if the state is –1, then the state is the majority of itself and its nearest neighbors to the south and west. (The ES/WN, SW/NE, and WN/ES directions perform identically, of course.) The DMA is defined to successfully "identify the majority" if (a) there *is* a majority of one of the states over the other at $t = 0$, and (b), after repeated iteration, it produces a uniform configuration consisting only of the initial majority state.

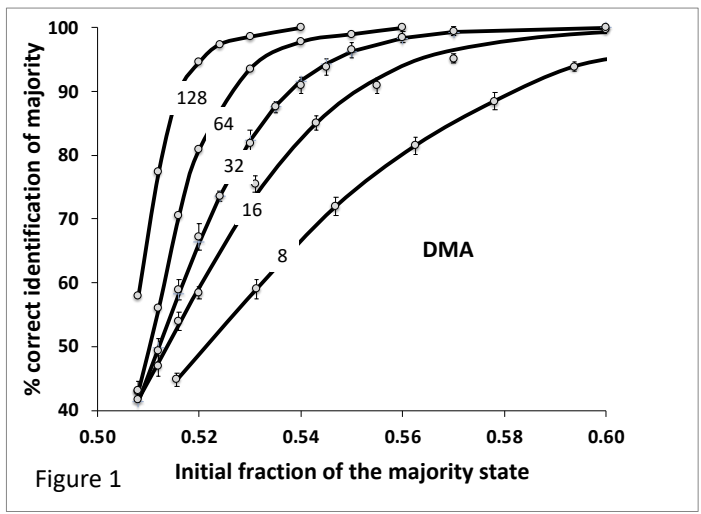

Figure 1

Figure 1 shows successful identification rates for the DMA as functions of the fractions of the initial majority states and for different space sizes (with $N = 8, 16, 32, 64,$ and $128$). Each data point shown is the average of $10^4$ simulations for randomly chosen initial configurations. As is the case for all binary cellular automata [Land (6), Bušić (7)], the DMA does not successfully identify the majority for *all* initial configurations. When the DMA *fails* to identify the initial majority state, the end configurations are either all of the wrong state, or unresolving mixtures of +1 and −1. As shown in Figure 1, the DMA is increasingly successful as the size of the space increases.

**The DMA Renormalization Flow**

The successful majority-identification curves in Figure 1 are observed to be well fit by the form

$$\sigma(N,f) = 100/(1 + \exp[\{AN + B\}\{2f - 1\} + \ln\{CN + D\}]), \quad (1)$$

where, $\sigma(N,f)$ is the percentage of the cases the DMA successfully identifies the majority initially present with fraction $f$, in a space of size $N$x$N$. In (1), for the data in Figure 1, $A = -1.006$, $B = -9.532$, $C = 0.042$, and $D = 1.433$. The uncertainty in these values is about $\pm 3\%$. Note that the form of (1) is also suggested by the observation that the DMA dynamics is analogous to a classical thermal system evolving to an equilibrium condition from an unstable initial state. In this analogy, the denominator in (1) can be interpreted as a grand canonical partition function, with $2f - 1$ playing the role of inverse temperature, $-(AN + B)$ the energy level of the system of size $N$x$N$, and $\ln(CN + D)$ the analog of (a temperature dependent) chemical potential.

The $2^n$x$2^n$ spaces referred to in Figure 1 have no majority when $f = 0.5$. For this case, the final configurations are either all +1, all −1, or periodically changing configurations of fixed numbers of +1 and −1. The value, $\sigma(N, 0.5)$, is the percent of the time a 50-50 configuration converges to all +1, or equally to all −1. The fractions of the time each of the +1, −1, and mixed configurations appear are determined by the size of the space—with the fraction of the mixed configurations increasing as $N$ increases (ranging from 28% for $N = 8$ to 74% for $N = 128$). When $f$ is increased from 0.5 by even a small amount, however, the DMA converges to the majority identification values shown in Figure 1. Thus, for the DMA, $f = 0.5$ can be interpreted as producing a critical condition of the dynamics.

Complex dynamical systems can sometimes be transformed into simpler equivalent forms guided by the system's "beta function." For the DMA it is useful to define such a function, $\tilde{\beta}(N,f)$, as

$$\tilde{\beta}(N,f) \equiv \frac{\partial \sigma}{\partial [\ln(f)]} = -2f\sigma^2(AN + B)(1/\sigma - 1). \quad (2)$$

Substituting the values for $A$ and $B$ from above shows that, for the DMA, $\tilde{\beta}(N,f) \geq 0$. At the critical value, $f = 0.5$, $\tilde{\beta}$ is positive and increases as $N$ increases: e.g., for $N = 8$, $\tilde{\beta} = 4.7$; $N = 16$, $\tilde{\beta} = 7.4$; $N = 32$, $\tilde{\beta} = 11.9$; $N = 64$, $\tilde{\beta} = 18.1$; $N = 128$, $\tilde{\beta} = 25.0$. Configurations with $f = 0.5$ are unstable and increasingly more so (as signaled by the increasing value of $\tilde{\beta}$) as the automaton space size increases. On the other hand, for all $N$, as $f$ approaches 1, $\tilde{\beta}$ approaches 0. Thus, *for all space sizes*, the DMA dynamics can be envisioned as "flowing" away from the critical configurations toward configurations of increasing $f$. In such a situation, the DMA dynamics can be "renormalized," that is, converted into another dynamical system with similar results, operating at coarser length scales [Wilson (8)].

**The DMA Universality Class**

To investigate consequences of the possible renormalization of the DMA, we examined the DMA's "critical exponents." As shown in Figure 1, small increases in $f$ above 0.5 lead to increased successful majority identification via the DMA. The initial majority fraction, $f$, can therefore be understood as a control parameter for the dynamics and the successful majority identification fraction, $\sigma$, can be viewed as a related order parameter. In the study of universal phenomena, the *critical exponent $\beta$* (*not* the value of the system's beta function) is defined through the relation

$$\frac{\sigma - \sigma_C}{\sigma_C} = A \left(\frac{f - f_C}{f_C}\right)^\beta, \text{ as } f \to f_C. \tag{3}$$

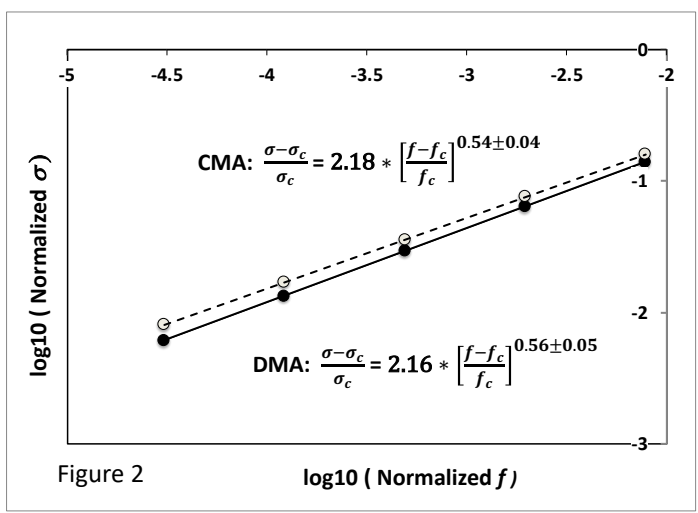

Figure 2

To evaluate $\beta$ for the DMA we approximated the limit in (3) using $f_c = 1/2$ and the values $f_1 > 1/2$ corresponding to one more +1 state than 50-50 in each configuration. For each value of $N$ we averaged $10^4$ simulations for random initial configurations with the same $f_c = 0.5$ and the same smallest $f_1$. The results are the black data points on the log-log plot shown in Figure 2.

As each data point is supposed to have the same $\beta$ value, the slope of the best fit line to the log-log data yields a good estimate of this parameter for all space sizes. From the data in Figure 2, we find $\beta = 0.56 \pm 0.05$, for the DMA. This result suggests that **DMA is arguably a member of the universality class of *directed percolation in 2 spatial dimensions***, i.e., for which $\beta = 0.583 \pm 0.003$ [Wang (9)].

**Coarsening the DMA**

Spatio-temporal dynamical systems that exhibit critical behavior are typically characterized by self-similar configurations of states. In such cases, assigning an effective state to blocks of the original configuration containing multiple cells, then running effective dynamics on the new values, can produce quantitatively similar results to the original dynamics, while greatly reducing the associated computational costs. We applied such a coarsening strategy—which we designate as the *Coarsened Majority Automaton*, the CMA—to the DMA dynamics.

Because the DMA always converges to all +1, all −1, or mixed +1s and −1s, we allowed our coarsening algorithm to incorporate *three* possible values: +1, −1, or 0. The coarsening rule we used is: replace every $2\text{x}2$ block with a single block and assign to it the state that was in the majority in the $2\text{x}2$; if there was no majority, assign to the block the state 0. Following that, we apply the dynamical rule: (a) if the block state is +1, set the new state to the majority of the block, the block to its north, and the block to its east; if there is no majority, set the new state to 0; (b) if the block state is −1, set the new state to the majority of the block, the block to its south, and the block to its west; if there is no majority, set the new state to 0; (c) if the block state is 0, set the new state to the majority of the block plus its four nearest neighbors. The blocks in the CMA are toroidally wrapped in the same manner as the cells in the DMA spaces. As the DMA in this study is defined on $2^n\text{x}2^n$ spaces, $n$ repetitions of this *coarsening+dynamics* process lead to a single block; for the CMA to successfully perform the majority task, then, the state of its final single block has to be +1 (or −1), whenever +1 (or −1) was in the majority in the initial configuration.

**Majority Identification by the Coarsened Algorithm**

It's not immediately obvious that the procedure outlined above would produce any relation to the original DMA. The CMA is, after all, a *different* automaton. As a first check to see if they might be related, we ran the CMA for $10^4$ random initial configurations with the same space sizes and values of $f$ as we did for the DMA. The results are shown in Figure 3.

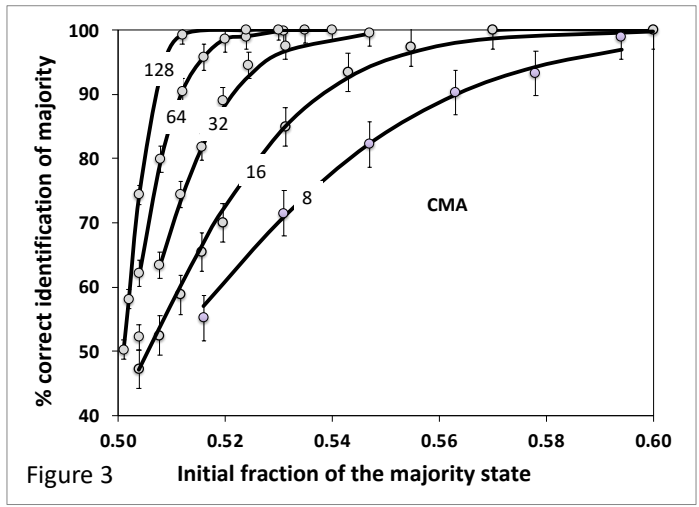

Figure 3

The data in Figure 3 are fit by the same functional form as Equation (1), but now with the values $A = -1.700$, $B = -6.623$, $C = 0.003$, and $D = 1.414$, the uncertainties of which are about $\pm 6\%$.

To further demonstrate that the DMA and the CMA are closely related we followed the same procedure as above for finding the CMA's critical exponent $\beta$. *The results are shown in the log-log plot of Figure 2 as the open circle data points.* The best fit to the CMA data yields a slope equal to $\beta = 0.54 \pm 0.04$, essentially indistinguishable from the result for the DMA. Thus, **the CMA is also likely to belong to the universality class of 2d directed percolation**.

It is natural to wonder if the DMA and CMA perform the initial majority identification task identically, or if one is better than the other. To investigate this question, we reran the CMA dynamics using each of the initial configurations as in Figure 1. Again, for each space size and for each value of the initial majority we ran and averaged the results of $10^4$ simulations. We found that the **CMA actually *outperforms* the DMA** using exactly the same initial configurations.

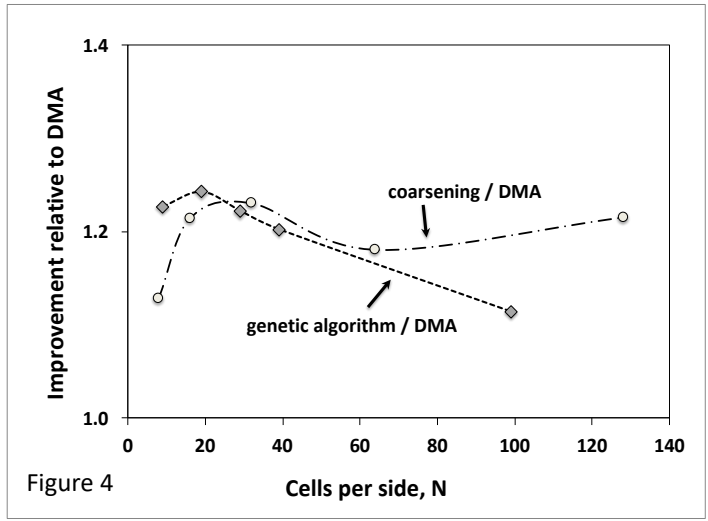

Figure 4

The results are summarized in Figure 4. The identification rate of the CMA is about 20% higher than that of the DMA when averaged over the same initial majority fractions for the different space sizes we used. Moreover, the CMA performs the task *much more economically*. For example, for 128x128 spaces, we observe that the DMA typically requires about 350 complete updates of all 16384 cell states to achieve a fixed or periodic end configuration— i.e., nearly $6 \times 10^6$ updates. For the same initial conditions, the CMA achieves its end state in *seven* iterations—involving a total of 5397 complete state updates, a factor of 1000 fewer computations.

Also shown in Figure 4 are results reported in reference 5. In that study, the authors employed a genetic algorithm search for alternative rules to the DMA to improve majority identification. The CMA's improvement over the DMA is at least equivalent to, and—for the data available in reference 5 and reported here—perhaps better than that obtained from the genetic algorithm search. And again, the CMA does so while very significantly reducing the associated computational cost compared to the best genetic algorithm found rule.

**Undecidability and Predictive Features in the DMA**

Figure 5 shows three 16x16 initial configurations with the same initial fraction of +1 (black) (52.3%) and –1 (white) (47.7%) cells. In the column to the right of the initial configurations are single blocks representing the 16x16 final configurations produced by the DMA dynamics. In the top row, the final configuration is 100% black. According to Figure 1, all black is produced about 60% of the time for this setup. In the middle row, the final configuration is 100% white. At the bottom, the final configuration–represented as gray–is an unresolving mixture of black (59.8%) and white (40.2%). If the DMA were a perfect majority identifier all three final configurations would be 100% black.

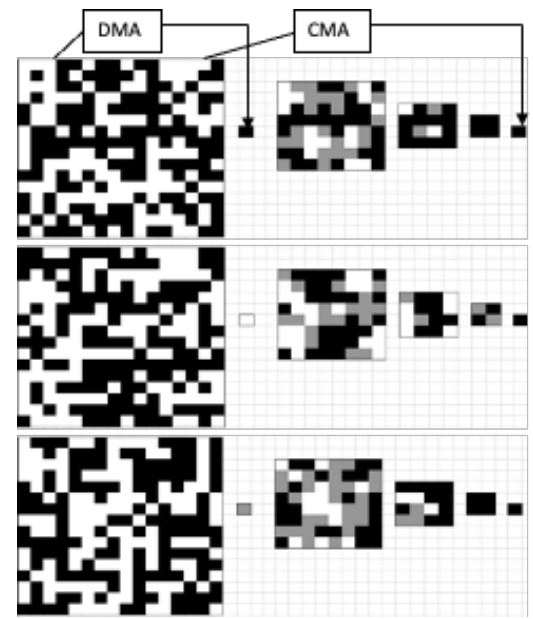

Figure 5

To the eye, the initial configurations shown in Figure 5 appear *qualitatively* identical, yet the final configurations are clearly different. In other words, the exact relation of all of the states to one another is obviously important for the DMA's output. This fussy dependence on the initial placement of states is referred to as "undecidability." In general, if an automaton is undecidable the only way—almost—to determine what it will produce for each initial configuration requires running the dynamics to near completion. [See, e.g., Wolfram (10)]

The reason the DMA is *almost* undecidable is that the final configuration is *observed* to almost always be preceded by the formation of one or more "percolation paths." A percolation path on a torus is a single state, nearest-neighbor connected path that *spans* the torus. It is closed, in the sense that a cell on it in the upper row is in the same column as a cell on it in the lower row (or in the same row as cells in the first and last columns). When the final configuration of a DMA run consists of all the same state, the percolation paths that form early on consist only of that final state. Such percolation paths catalyze a system-wide collective state change. In other words, **successful identification of the initial majority by the DMA is equivalent to a directed percolation phase transition**.

When the DMA evolves to a mixed configuration there will either be two nonintersecting percolation paths of opposite state forming at different times during the intermediate dynamics, or no percolation paths at all. Typically, such predictive precursor structures—one or more paths of the same state or nonintersecting paths of both states—emerge in relatively few of the time steps necessary to reach the corresponding final state. (The rare no-percolation-paths case takes longer.) Thus, the DMA is *almost* undecidable.

While it is generally still impossible to infer the outcome of a DMA process by just examining its initial configuration (unless there is already a percolation path present, such as in the top initial configuration in Figure 5), the early formation of percolation paths after only a small number of iterations of the dynamics provides useful clues.

**Why Does the CMA Outperform the DMA in the Majority Identification Task?**

Figure 5 also shows the results of the CMA dynamics evolving the same initial configurations as given to the DMA. Starting with 16x16 states, there are in each case successively 8x8, 4x4, and 2x2 states, and finally one single state. The CMA states can be black (+1), white (–1), or gray (0). In each of the cases shown, the final all black configuration correctly identifies the initial majority.

In each of these cases, the successively coarser configurations contain increasing fractions of the initial majority state. For the purpose of calculating the black content fraction, gray can usefully be interpreted as half black, half white. In the top trial depicted in Figure 5, the black content fraction for the CMA increases from 52% (for 16x16) to 59% (for 8x8) to 81% (for 4x4) to 100% (for 2x2). In the middle trial the black fractions are 55%, 56%, and 75%. In the bottom trial the fractions are 58%, 75%, and 100%. This sequential majority amplification appears to be a general result. For example, an average of 1000 experiments starting each time with the same initial black fraction (52.3%) as in Figure 5 *and* where the CMA converges to one black cell (which happens 78% of the time), yields the increasing sequence, 58.5±0.4% (for 8x8), 74.3±1.0% (for 4x4), and 95.7±1.4% (for 2x2). *In other words, whenever the coarsening algorithm successfully identifies the initial majority, it always amplifies the initial state fraction in each coarsening.* (This conclusion isn't surprising: if at one coarsening stage the fractions of states switched, then that stage would set the initial conditions for subsequent coarsenings and consequently jeopardize the identification of the majority.)

In addition to state amplification, whenever the coarsening algorithm correctly identifies the initial majority, at least one percolation path consisting of all the majority state emerges in the coarsened configurations before the final configuration. In the examples in Figure 5, after each first coarsening (i.e., the 8x8 configurations) there is at least one black percolation path—counting gray as 50% black—but no white one.

Therefore, the reason the CMA outperforms the DMA is because when the coarsening algorithm eventually identifies the initial majority it increasingly amplifies the initial state fraction in each coarsening step while at the same time sequentially involving *fewer and fewer cells*. It is more likely, therefore, that the CMA will produce closed percolation paths of the correct state—*the progenitors of correct identification*—than the DMA.

**Summary**

The Directed Majority Automaton is a two-dimensional, 2-state cellular automaton that competently, though not perfectly, performs the initial majority task over a wide range of initial majority fractions. This feat is accomplished autonomously by collective dynamics. Here, we demonstrate that the collective dynamics of the DMA resides in the universality class of directed percolation. We observe that successful majority identification is always preceded by the emergence of a closed percolation path consisting solely of the initial majority state. Such paths subsequently initiate a phase transition in which all of the cells of the automaton take on the initial majority state. Exploiting the possibility that the collective dynamics of the DMA might be renormalizable, we develop a coarsened version of the DMA (the CMA) which resides in the same universality class as the DMA, and which performs the majority identification task by forming percolation paths in the coarsened state spaces. The CMA accomplishes this with higher success rates and vastly fewer computations than the DMA.